\documentclass[twocolumn]{svjour3}
\journalname{epja}
\usepackage[utf8]{inputenc}
\usepackage[pdftex, colorlinks=true, linkcolor=blue, urlcolor=blue, citecolor=blue]{hyperref}
\usepackage{glossaries-extra}
\setabbreviationstyle{long-short}
\glsdisablehyper
\newabbreviation[longplural={Equations of State}]{eos}{EoS}{equation of state}
\newabbreviation[longplural={degrees of freedom}]{dof}{dof}{degree of freedom}
\newabbreviation[shortplural={HIC},longplural={Heavy-Ion Collisions}]{hic}{HIC}{heavy-ion collision}
\newabbreviation[description={Nuclotron-based Ion Collider fAcility in Dubna, Russia}]{nica}{NICA}{Nuclotron-based Ion Collider fAcility}
\newabbreviation{rdft}{RDFT}{relativistic density functional theory}
\newabbreviation{rdf}{RDF}{relativistic density functional}
\newabbreviation{sfm}{SFM}{string-flip model}
\newabbreviation{qcd}{QCD}{quantum chromodynamics}
\newabbreviation{qgp}{QGP}{quark-gluon plasma}
\newabbreviation{njl}{NJL}{Nambu-Jona-Lasinio}
\newabbreviation{nse}{NSE}{nuclear statistical equilibrium}
\newabbreviation{qs}{QS}{quantum statistical}
\newabbreviation{cve}{CVE}{cluster virial expansion}

\newglossaryentry{cern}{name={CERN},description={European Organization for Nuclear Research in Geneva, Switzerland}}
\newglossaryentry{lhc}{name={LHC},description={Large Hadron Collider Experiment at \gls{cern}}}
\newglossaryentry{alice}{name={ALICE},description={A Large Ion Collider Experiment}}
\newglossaryentry{rhic}{name={RHIC},description={Relativistic Heavy Ion Collider at the Brookhaven National Laboratory in Upton, USA}}
\newglossaryentry{fair}{name={FAIR},description={Facility for Antiproton and Ion Research at GSI in Darmstadt, Germany}}
\newglossaryentry{dd2all}{name={DD2(F)},description={Walecka-type hadron model with density dependent couplings}}
	\newglossaryentry{dd2}{name={DD2},parent=dd2all,description={}}
	\newglossaryentry{dd2f}{name={DD2F},parent=dd2all,description={}}
\newglossaryentry{qmd}{name={QMD},description={quantum molecular dynamic}}
\newglossaryentry{amd}{name={AMD},description={AMD}}
\newglossaryentry{theseus}{name={THESEUS},description={THESEUS}}
\newglossaryentry{urqmd}{name={UrQMD},description={UrQMD}}
\newglossaryentry{rhic:bes}{name={RHIC BES},description={BES at RHIC}}
\newglossaryentry{rhic:fxt}{name={RHIC FXT},description={FXT at RHIC}}
\newglossaryentry{na61}{name={NA61},description={NA61}}
\newglossaryentry{lattice_qcd}{name={Lattice QCD},description={Lattice QCD}}

\usepackage{subfigure}

\usepackage{graphics,graphicx}

\usepackage{amsmath, amssymb}
\usepackage{mathtools}
\usepackage{multirow}
\usepackage{longtable}
\usepackage{color}
\usepackage{slashed}
\usepackage[export]{adjustbox}

\usepackage{cleveref}
\crefname{equation}{Eq.}{Eqs.}
\crefname{figure}{Fig.}{Figs.}
\crefname{tabular}{Tab.}{Tabs.}

\usepackage[autolanguage,np]{numprint}
\renewcommand*\npstyleenglish{\npthousandsep{~}\npdecimalsign{.}\npproductsign{\cdot}\npunitseparator{~}\npdegreeseparator{}\npcelsiusseparator{\nprt@unitsep}\nppercentseparator{\nprt@unitsep}}
\newcommand{\agrscale}{0.38}
\newcommand{\diascale}{0.12}

\usepackage[normalem]{ulem}  
\usepackage{subfigure}

\let\oldhref\href
\renewcommand{\href}[2]{\oldhref{#1}{\hbox{#2}}}

\begin{document}

\title{%
A unified quark-nuclear matter equation of state from the cluster virial expansion within the generalized Beth-Uhlenbeck approach%
}
\titlerunning{A unified quark-nuclear matter EoS}

\author{Niels-Uwe Friedrich Bastian \and David Bernhard Blaschke}

\institute{Niels-Uwe Friedrich Bastian \and David Bernhard Blaschke
	\at		Institute of Theoretical Physics, University of Wroclaw, 50-204 Wroclaw, Poland
	\and	David Bernhard Blaschke 
	\at		Bogoliubov Laboratory for Theoretical Physics, JINR Dubna, 141980 Dubna, Russia
	\and	National Research Nuclear University (MEPhI), 115409 Moscow, Russia
}

\date{Received: date / Accepted: date}

\maketitle

\begin{abstract}
We consider a cluster expansion for strongly correlated quark matter where the clusters are baryons with spectral properties that are described within the generalized Beth-Uhlenbeck approach by a medium dependent phase shift.
We employ a simple ansatz for the phase shift which describes an on-shell bound state with an effective mass and models the continuum by an anti-bound state located at the mass of the three-quark continuum threshold, so that the Levinson theorem is fulfilled by construction.
The quark and baryon interactions are accounted for by the coupling to scalar and vector meson mean fields modelled by density functionals.
At increasing density and temperature, due to the different medium-dependence of quark and baryon masses, the Mott dissociation of baryons occurs and its contributions to the thermodynamics vanish. 
It is demonstrated on this simple example that this unified approach to quark-hadron matter is capable of describing crossover as well as first order phase transition behaviour in the phase diagram with a critical endpoint.  
Changing the meson mean field, the case of a ``crossover all over'' in the phase diagram is also obtained. 
\end{abstract}

\section{Introduction}
\label{sec:intro}
The investigation of the phase diagram of \gls{qcd} is one of the major goals of nuclear and particle physics. 
Monte-Carlo simulations of the \gls{qcd} partition function on space-time lattices have reached a precision stage where discretization errors and the continuum limit are under control and calculations are performed for physical quark masses providing the pseudocritical temperature of the hadron-to-quark matter crossover transition at vanishing baryo-chemical potential $\mu$ as
$T_\mathrm c(\mu=0)=156.5\pm 1.5$ MeV \cite{Bazavov:2018mes}.
Unfortunately, these calculations face the sign problem for 
$\mu>0$ and cannot elucidate the \gls{qcd} phase structure in the whole $T-\mu$ plane where a central aim is to identify the position of one (or several) critical endpoint(s) beyond that of the nuclear gas-liquid transition. 
Perhaps there is no such critical endpoint and the hadron-to-quark matter transition is a crossover all over  the \gls{qcd} phase diagram. 

In order to make reliable predictions for the QCD phase structure, effective non-perturbative approaches to low-energy QCD at finite $T$ and $\mu$ have been developed, such as the Dyson-Schwinger equation approach \cite{Roberts:2000aa}.
Within this approach remarkable progress has been achieved towards a theory of the QCD phase diagram and a unified \gls{eos} of quark-hadron matter \cite{Fischer:2018sdj}. However, the inclusion of baryons into this scheme is still an open task.
Another systematic non-perturbative approach to low-energy QCD is based on applying the functional renormalization group methods. For an overview, see \cite{Pawlowski:2005xe}.
On the basis of the Polyakov-Quark-Meson model, the phase diagram with a critical endpoint could be obtained, albeit without baryons \cite{Schaefer:2007pw,Steinheimer:2013xxa}.
Recently, this approach has been developed towards a formulation of the hadronization problem, i.e. to describe hadrons as bound states of quarks \cite{Alkofer:2018guy}.

While these fundamental approaches are being further developed it is worthwhile to construct effective dynamical models for QCD thermodynamics which are more practical to use and at the same time provide a scheme for the unified description of hadronic and quark-gluon degrees of freedom where hadrons appear as composites (clusters) of quarks and gluons. 

In the description of the quark-hadron matter transition it became customary to employ the so-called two-phase approaches where quark and hadronic matter EoS are modelled separately and subsequently joined by a phase transition construction as, e.g., the Maxwell construction.
For an early review on such constructions, see \cite{Cleymans:1985wb}. A more recent overview on the thermodynamics of the deconfinement transition with relevant references is given in \cite{Satz:2012zza}. 
This approach has the disadvantage that it ignores the fact that hadrons are bound states of quarks and their thermodynamics should be directly related to the underlying quark dynamics \cite{Bastian:2015avq}.
By construction, you will get always a first-order phase transition, what is a contradiction to the Lattice results for low chemical potentials.
A van-der-Waals model for hadrons allows the construction of a critical endpoint end hence a cross-over at high temperatures \cite{Vovchenko:2017gkg}.
Similar results one obtains with density and temperature dependent excluded volume approaches \cite{Typel:2017vif}.
These transitions are purely thermodynamic and therefore do not result in a transition to quark matter.
An alternative approach is using a switching function to match a hadronic and quark \gls{eos} \cite{Albright:2014gva,Parotto:2018pwx}.

The goal would be to describe the hadrons as solutions of the equations of motion of multi-quark states in medium so that their dissociation within the Mott effect is obtained naturally.
The quark interactions shall capture the most important aspects of confinement, namely, that they allow a sufficient number of hadronic eigenstates up to masses of, e.g., $m_\mathrm{max} \approx \np[GeV]{1.5}$, and that no free quark states are excited before the chiral transition is reached.
An approach based on a Hagedorn bag-like model is described in \cite{Vovchenko:2019jyl}.

The simultaneous description of the thermodynamics of bound states (clusters) and their constituents is the subject of the physics of non-ideal plasmas as well as the particle clustering and Mott transitions in nuclear matter \cite{Ropke:1982ino,Ropke:1983lbc}. 
This description has recently been advanced to a cluster virial expansion \cite{Ropke:2012qv} on the level of the generalized Beth-Uhlenbeck approach for nuclear matter
\cite{SCHMIDT199057}.
The extension of such a description to quark matter systems (where the elementary degrees of freedom are the quarks and the clusters are the hadrons) has been worked out in Ref.~\cite{Blaschke:2013zaa} (see also references therein).
It meets the problem that quark confinement at low densities has to be taken into account. 
This can effectively be achieved by coupling the chiral quark dynamics to the Polyakov loop \cite{Ratti:2005jh,Roessner:2006xn}.
However, at low temperatures and finite baryon densities such a description lacks the knowledge of the Polyakov-loop potential, because no lattice QCD calculations exist in that domain for fitting it.
An alternative has been suggested in the form of a relativistic density functional approach that suppresses quarks at low densities by a diverging scalar self-energy (mass term), see Ref.~\cite{Kaltenborn:2017hus}. 

In the present work we will employ the concept, which is described in \cite{Bastian:2018wfl}, for the implementation of quark confinement/deconfinement and describe the appearance/disappearance of baryons within a simplified realization of the generalized Beth-Uhlenbeck approach that works with generic ansatz for the medium dependence of the baryon phase shifts that encodes their Mott dissociation.
As a result we shall obtain the thermodynamics of the non-ideal quark-baryon plasma in the form of a cluster expansion and will be able to relate the existence of a critical endpoint of a first order deconfinement transition to the microphysics of the model.
Within the scope of the current study we restrict ourselves to iso-spin symmetric matter.

\section{Cluster Expansion}
\label{sec:cve}

\begin{figure*}[tb]
	{\centering
    \hfill
    \hfill
    $\underbrace{
	\includegraphics[scale=\diascale,valign=t]{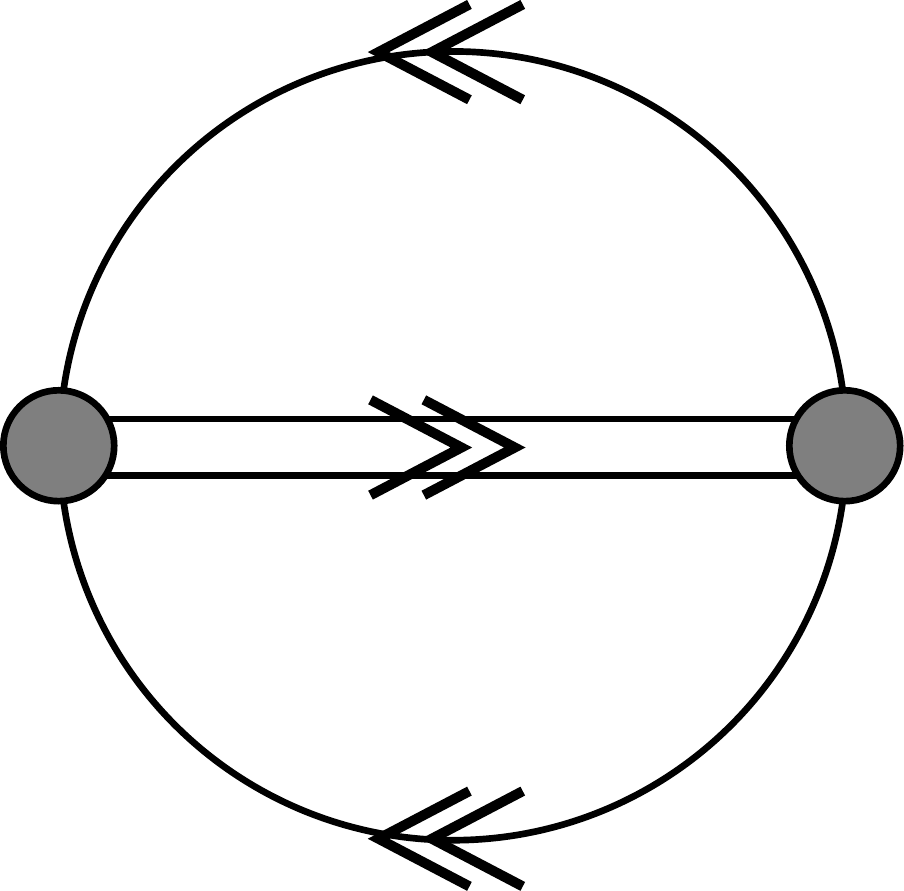}%
    \parbox{5mm}{\vspace{1.5cm} \Large $+$}
	\includegraphics[scale=\diascale,valign=t]{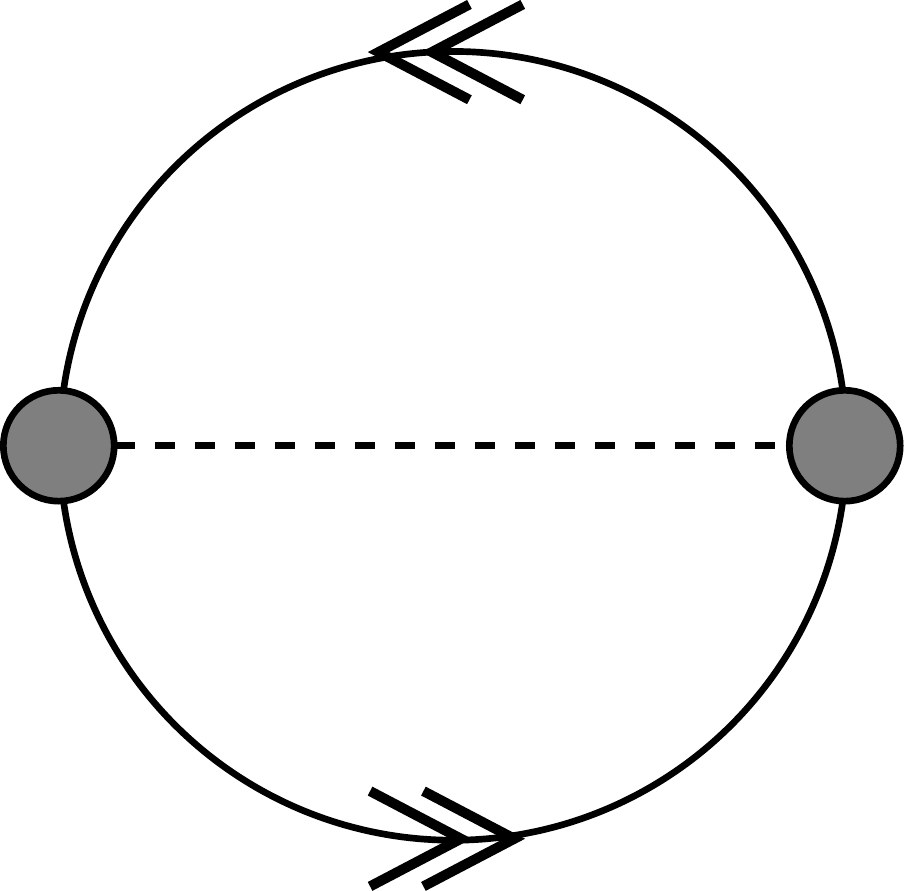}%
    }_{	
    \includegraphics[scale=\diascale,valign=t]{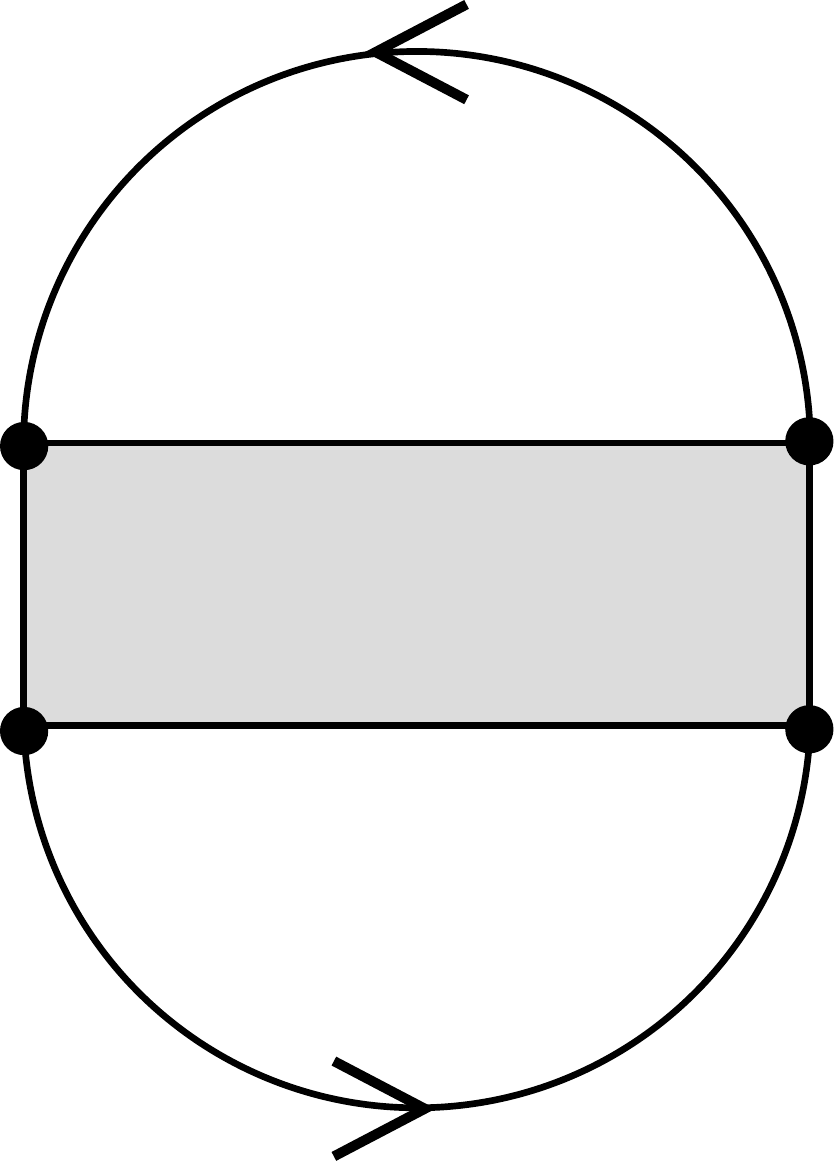}%
	}$
    \hfill
    \parbox{5mm}{\vspace{1.5cm} \Large $+$}
    \hfill
    $\underbrace{
	\includegraphics[scale=\diascale,valign=t]{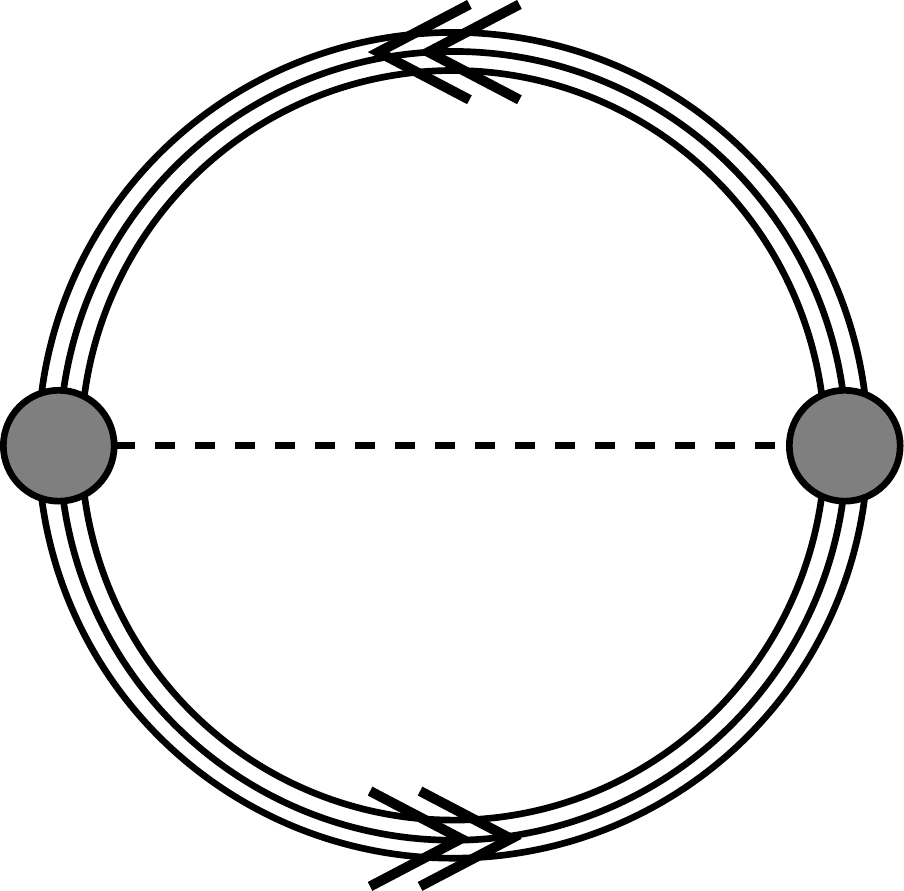}%
    }_{	
    \includegraphics[scale=\diascale,valign=t]{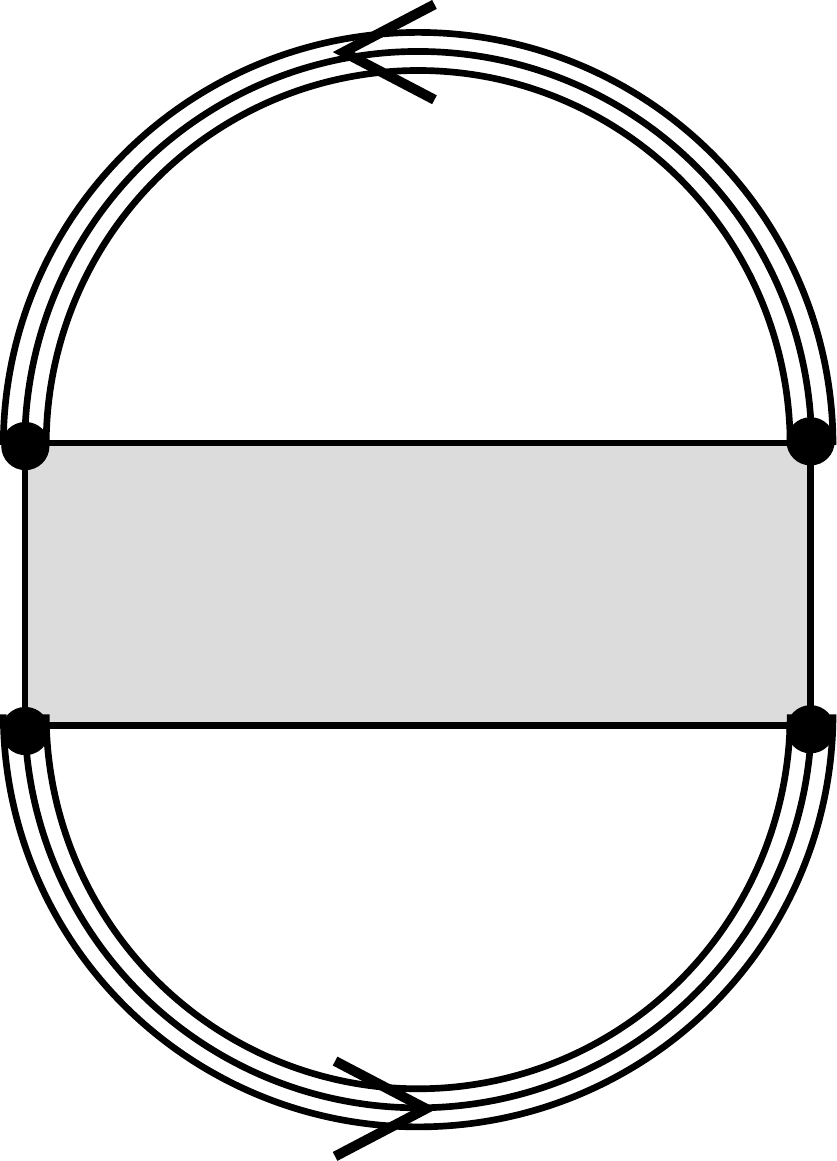}%
	}$
    \hfill
    \parbox{5mm}{\vspace{1.5cm} \Large $+$}
    \hfill
    $\underbrace{
	\includegraphics[scale=\diascale,valign=t]{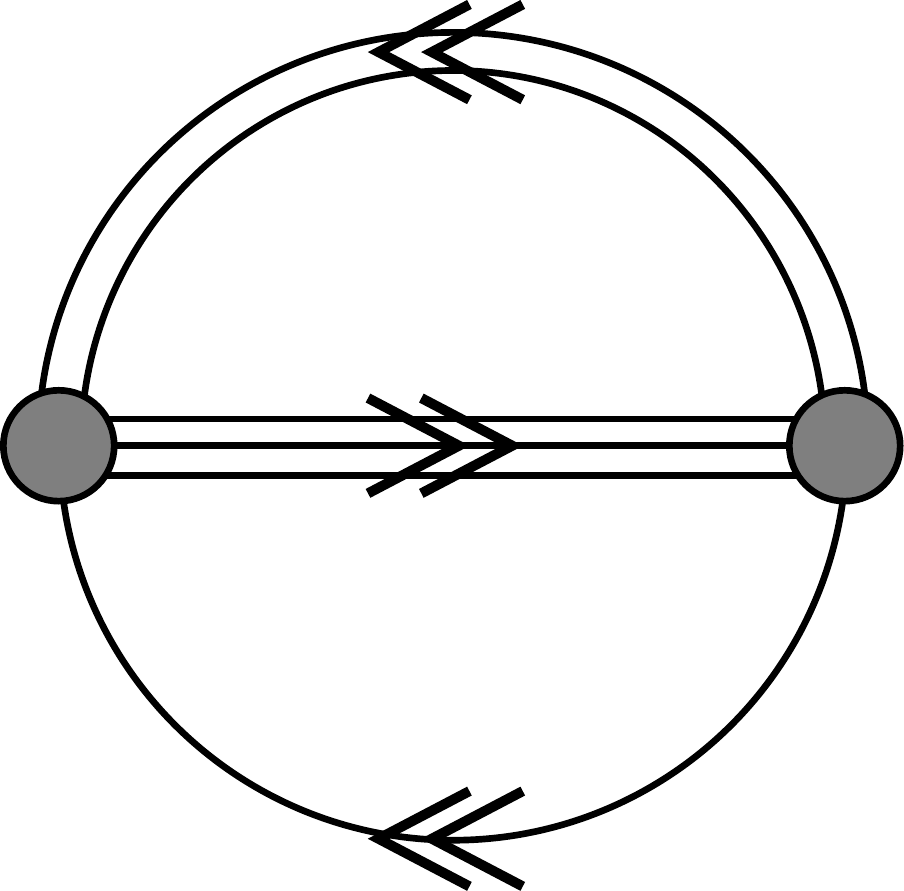}%
    }_{	
	\includegraphics[scale=\diascale,valign=t]{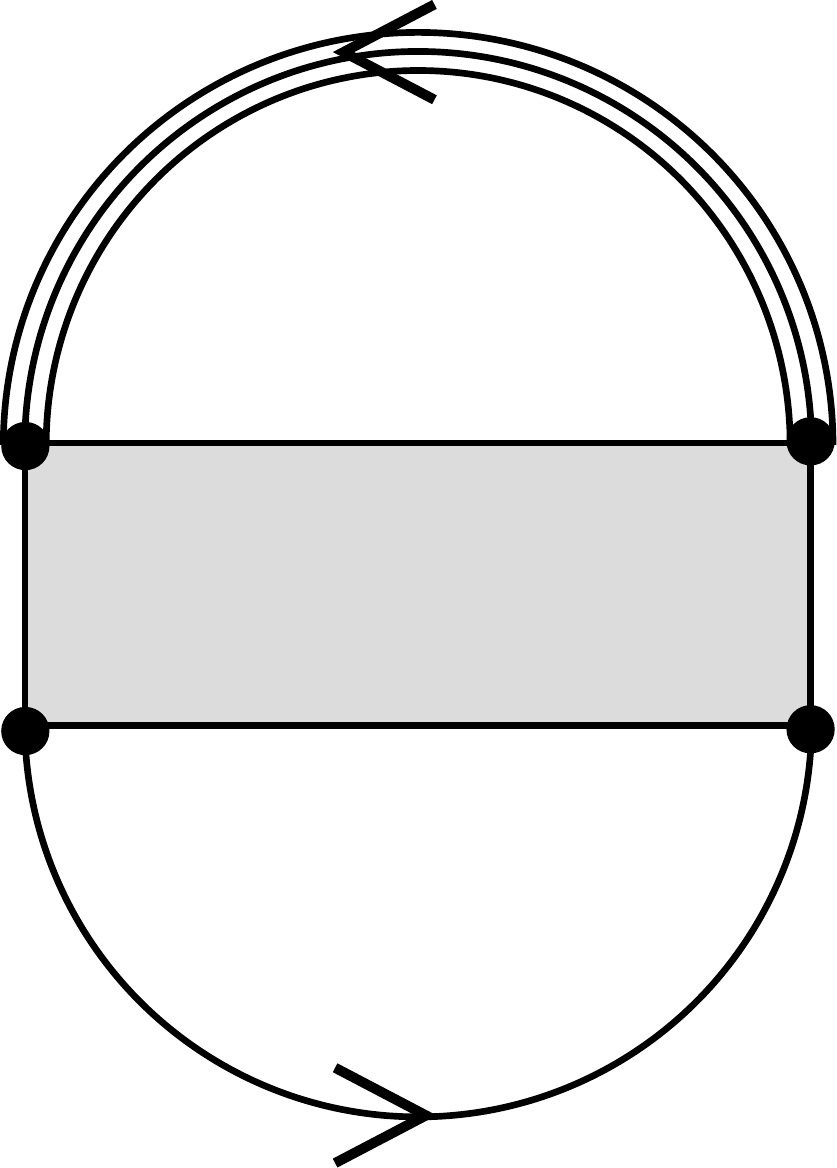}%
	}$
    \hfill
    \parbox{5mm}{\vspace{1.5cm} \Large $+$}
    \hfill
    $\underbrace{
	\includegraphics[scale=\diascale,valign=t]{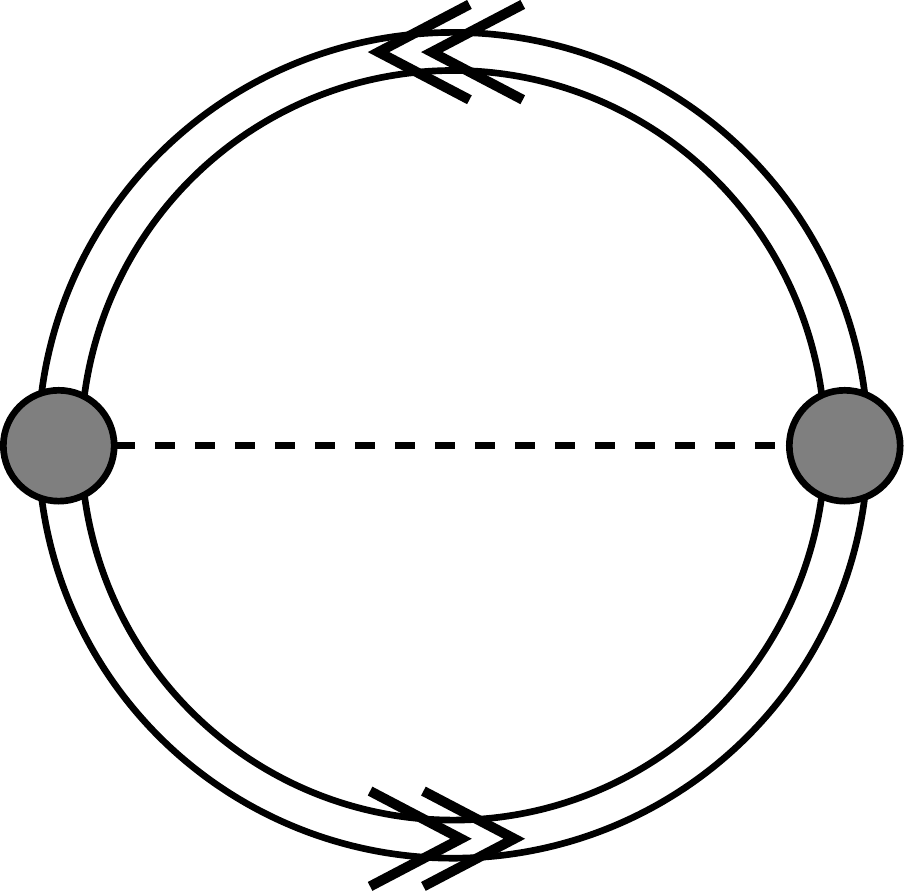}%
    }_{
    \parbox{5mm}{\vspace{1cm}\includegraphics[scale=\diascale,valign=t]{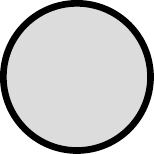}}%
    }$
    \hfill
    \hfill}
	\caption{%
    Contributions of the baryon-meson-quark-diquark system to the $\Phi$-functional.
    The upper row shows the full set of sunset type diagrams.
    The lower row shows the resulting diagrams to be taken into account, after collapsing the diquark and meson-propagators.
    The first and second diagram represent the mean field contributions to the quark and the baryon propagators.
    Here the contributions result from effective interactions with the meson and diquark fields.
    The third diagram represents the coupling of baryons to quarks, which is elementary done by diquarks, but here described as an effective coupling term.
    The last term, which is the remainder of the meson-diquark contributions is a constant of the thermodynamic potential, whose derivatives (like particle density) vanish.
    }	
	\label{fig:diagrams}
\end{figure*}

It is shown in \cite{Blaschke:2016fdh,Bastian:2018wfl}, that the $\Phi$--derivable approach \cite{Baym:1961zz,Baym:1962sx} can be generalized to resemble a cluster decomposition of the thermodynamic potential
\begin{align}\label{Phi-A}
\begin{aligned}
	\Omega = \sum_{i} \Omega_i = \sum_{i} \left\{\!\!\vphantom{\sum}\right. \frac{c_i}{2} \left[\operatorname{Tr} \ln \left(-G_i^{-1}\right) + \operatorname{Tr} \left(\Sigma_i~G_i \right) \right]\\
    + \sum_{\stackrel{j,k}{j+k=i}}\Phi[G_i,G_j,G_k]\left.\vphantom{\sum}\!\!\right\}\,,
\end{aligned}
\end{align}
where $i$ labels the particle species and stands for quark, meson, diquark and baryon contributions, $c_i=+1(-1)$ for bosons (fermions).
The full cluster Green's function $G_i$ fulfils the Dyson equation $G_i^{-1} = G_{i,0}^{-1} - \Sigma_i$, where the cluster self energy $\Sigma_i$ is defined as the functional derivative of the $\Phi$--functional $\Sigma_i = {\delta \Phi}/{\delta G_i}$.
In the cluster virial expansion scheme that we follow, we restrict ourselves to the choice of all two-loop diagrams of the ``sunset'' type that can be drawn with three cluster Green's functions for cluster sizes $i$, $j$ and $k$ that fulfil the relation $i=j+k$ with $i>j, k$ in order to preserve charge conservation at the vertices.
For the description of the system within this approach, one would need to include the diagrams, shown in the upper row of \cref{fig:diagrams}, as contributions to the $\Phi$-functional \cite{Bastian:2018wfl}.
Before solving the entire system, we want to do a first step, by replacing the meson and diquark propagators by their mean fields, as it is also shown in the lower row of diagrams in \cref{fig:diagrams}.
Here we neglect the meson-meson couplings, which would lead to purely mesonic diagrams.
These types of vertices would only play a role in the non-linear extension of the model, which is not considered at this point.
The remaining particles, baryons and quarks, are affected by a mean field, coming from the diquark and meson contributions, which we describe effectively by density-dependent self-energies.
The third term describes the coupling of quark and baryon sectors and the last term is a remnant of the meson-diquark interaction which in our approximation becomes a constant of the thermodynamic potential and therefore its derivatives, such as particle density, 
vanish.
As it is shown in \cite{Vanderheyden:1998ph,Blaizot:2000fc}, for both types of diagrams in \cref{fig:diagrams}, a cancellation holds which allows to write the cluster expansion in the form of the generalized Beth-Uhlenbeck approach \cite{Bastian:2018wfl,Blaschke:2016fdh}
\begin{align}
	n_j &= -\left(\frac{\partial \Omega}{\partial \mu_j}\right)_{T,\mu_{k\neq j}} = \sum_{i=1} A_{ij} n_i(T,\mu)\\
\begin{split}
    &= \sum_{i=1} A_{ij} g_i \int \frac{\mathrm d^3p\;\mathrm d E}{(2\pi)^4}f_i(E)  2 \sin^2 \delta_i(E) \frac{\mathrm d\delta_i (E)}{\mathrm d E}~,
\end{split}
\end{align}
with the phase shift $\delta_i (E)$ as a medium-dependent quantity, which already includes all properties of the cluster $i$, corresponding to its spectrum of bound and scattering states.
The matrix $A_{ij}$ is formed by the number of constituents $j$ in the cluster $i$, while $j \subset i$.
The species $i$ has degeneracy factor $g_i$ and obeys the Fermi distribution $f_i(E) = \left(\exp [(E-\mu_i)/T] + 1\right)^{-1}$.

The chosen diagrams in the lower row of \cref{fig:diagrams} are of Hartree type.
The resulting self energies thus are real and represent shifts of the poles of the quark and nucleon quasi-particle propagators.
Hence the phase shifts defined as $\delta_i = \arctan ({\operatorname{Im}G_i}/{\operatorname{Re}G_i})$, degenerate to functions which can only attain values of $n \pi$, where $n$ is an integer standing for the number of bound states.
Using the quasi-particle dispersion relation
\begin{align}\label{eq:energydisperion}
	E_\mathrm{i} &= \sqrt{p^2 + (m_i + S_i)^2} + V_i~,
\end{align}
with the scalar (vector) self energy $S_i$ ($V_i$), we can substitute the energy of the quasi particle by its effective mass $M_i = m_i + S_i$ and obtain
\begin{align}
\begin{split}
	n_{i}^\mathrm{free} &= g_i \int \frac{\mathrm d^3p}{(2\pi)^3} \int \frac{\mathrm d M}{2\pi} f_i\left(\sqrt{p^2 + M^2} + V_i\right) \times\\
		&\times 2 \sin^2 \delta_i(M) \frac{\mathrm d\delta_i (M)}{\mathrm d M}~.
\end{split}\label{eq:bu_density}
\end{align}
Here we replaced the energy-dependent phase shift $\delta_i (E)$ by a (effective) mass dependent representation $\delta_i (M)$, which absorbs the contributions of scalar self energy.

Within this letter, we consider isospin-symmetric matter and therefore can reduce the quark flavours to one degenerate one $j = \{\mathrm{q}\}$ as constituent particles and additionally nucleons (protons and neutrons) as degenerate composites so that $i = \{\mathrm{q},\mathrm{n}\}$.
Hence the degeneracy factors are $g_i = (12,4)^\mathrm T$ and quark constituent of the clusters is $A_{ij} = (1,3)^\mathrm T$.
The generalization to isospin-asymmetric or strange matter and the inclusion of other baryons (e.g. hyperons), can be done analogously.
We define the phase shifts as
\begin{align}
	\delta_{i = \mathrm{n}} (M) &= \pi \, \Theta (M - M_i) \, \Theta (M_i^\mathrm{thr} - M)~,\\
	\delta_{i = \mathrm{q}} (M) &= \pi \, \Theta (M - M_i)~,
\end{align}
for nucleons and quarks, respectively.
\begin{figure}[!ht]
	\centering
	\includegraphics[scale=0.3]{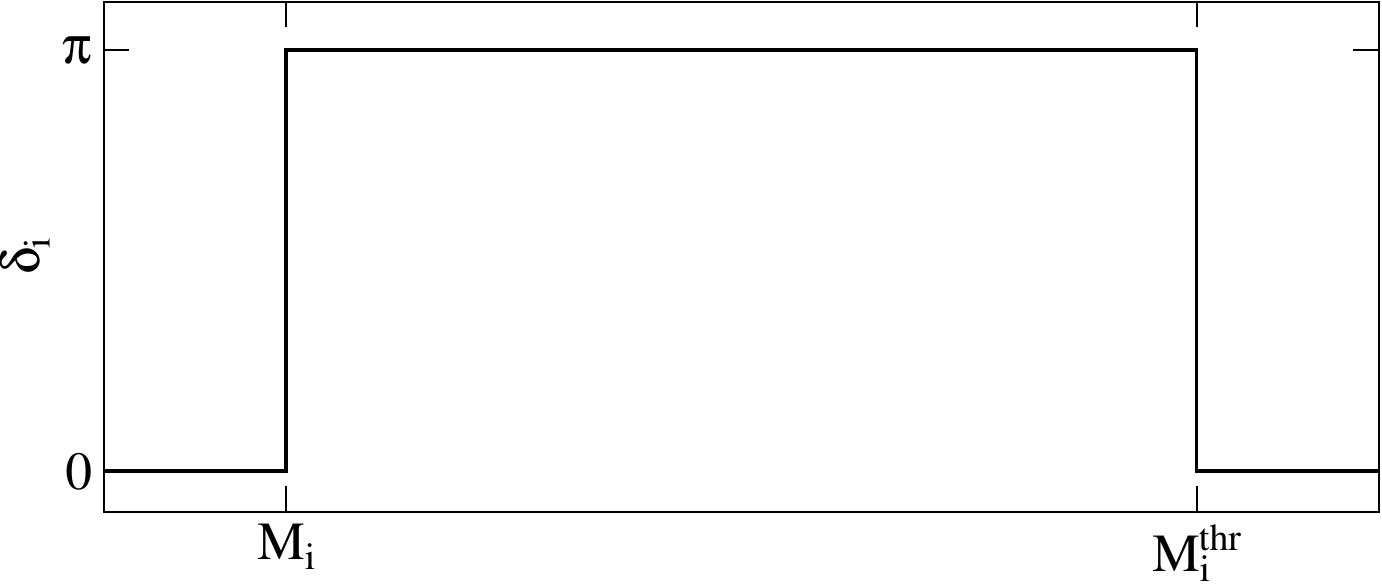}
	\caption{Phase shift of nucleons as function of mass. For both asymptotic limits $M = \pm \infty$ it vanishes, in agreement with the Levinson theorem. At the quasi-particle mass $M_i$ of the nucleon it has a sharp jump from zero to $\pi$, referring to the bound state. At the continuum threshold mass $M^\mathrm{thr}_i$ it jumps back from $\pi$ to zero, resembling a continuum phase shift that degenerates to an anti-bound state.}
	\label{fig:phaseshift}
\end{figure}
The simple ansatz for the phase shift as depicted in \cref{fig:phaseshift} has already been utilized before in the description of composite pions in quark matter in \cite{Blaschke:2017pvh}.
It features a bound state (jump up $0 \rightarrow \pi$) at the nucleon mass $M_i$ and an anti-bound state (jump down $\pi \rightarrow 0$) at the continuum threshold mass $M_i^\mathrm{thr}$, satisfying in this manner Levinson's theorem.
Here the spectral density of the continuum states degenerates to that of a negative delta distribution at the continuum threshold and acts like the negative of the contribution from a bound state in the same channel.
As elementary particles, quarks are here described as quasi particles without substructure.
The analogue of the threshold in this case is at positive infinity.

The threshold masses of nucleons are defined as the sum of the masses of their constituents $M_\mathrm n^\mathrm{thr} = 3 M_\mathrm q$.
In this way, the bound state is density-dependent via the self-energy shifts of the quarks as well as the nucleons and, once the constituent masses drop below the nucleon mass, the Mott transition occurs.

Due to the simple shape of the phase shifts, the mass integration in \cref{eq:bu_density} can be performed analytically.
Note here, that the derivative of the Heavyside distributions $\Theta(x)$ results in Dirac distributions and $\sin(\pi/2) = 1$ at the steps from zero to $\pi$.
As result we get the usual quasi particle density for (unbound) quarks
\begin{align}
	n_{i = \mathrm{q}} &= g_i \int \frac{\mathrm d^3p}{(2\pi)^3} f_i(\sqrt{p^2 + M_i^2} + V_i)
\end{align}
and an altered expression for (bound) nucleons
\begin{align}
\begin{split}
	n_{i = \mathrm{n}} &= g_i \int \frac{\mathrm d^3p}{(2\pi)^3} \left[ f_i(\sqrt{p^2 + M_i^2} + V_i)\right.\\
		&\left.- f_i(\sqrt{p^2 + (M_i^\mathrm{thr})^2} + V_i) \right] \Theta (M_i^\mathrm{thr} - M_i)
\end{split}\\
		&= \left(n_{\mathrm{N},i}^\mathrm{qu} - n_{\mathrm{N},i}^\mathrm{thr}\right) \Theta (M_i^\mathrm{thr} - M_i)~,
\end{align}
which contains one usual ``free'' contribution, subtracted by another contribution, based on the mass of the three-quark continuum threshold.
One can see that once the threshold mass drops below the nucleon mass at the Mott transition, the entire nucleon particle density vanishes.
In chemical equilibrium the chemical potentials $\mu_i$ of all particles $i$ can be expressed by their conserved charges.
For symmetric nuclear matter, where only the baryon chemical potential $\mu_\mathrm B$ is relevant, this leads to $\mu_i = B_i \mu_\mathrm B$, where $B_i$ is the baryon number of the particle $i$.

Given the relation between densities and chemical potentials, one can obtain the free energy density
\begin{align}
    f(T,n) &= \int_0^n \mu(T,n') \,\mathrm dn'\,.
\end{align}
Consequently, one can easily obtain the pressure as $p = \mu n - f$ and construct the phase transitions by fulfilling the Gibbs conditions of phase equilibrium.

\section{Self-energies}
\label{sec:selfenergies}
Recently, we developed a \gls{rdf} formalism \cite{Kaltenborn:2017hus}, which is capable of handling more complex types of interaction (such as confinement) on the mean field level.
Models within this formalism are defined by a density functional $U(n_\mathrm{s}, n_\mathrm v, n_\mathrm{vi})$, representing the interaction contributions.
Here $n_\mathrm{s}$, $n_\mathrm v$, $n_\mathrm{vi}$ are the scalar--isoscalar, vector--isoscalar and vector--isovector densities, which appear for quarks or nucleons as the expectation values of the corresponding bilinear expressions of quark and nucleon fields, respectively.
Based on the density functional, one can derive the scalar $S$ and vector $V$ self energies, which are used in the dispersion relation \cref{eq:energydisperion}.
As the scalar self energy defines the effective masses of particles, it represents the imperative element for the Mott transition.

\subsection{Quarks}
\label{sec:se:qu}
The model for confinement is motivated by the \gls{sfm} \cite{Ropke:1986qs}, and already achieved great success in the studies of astrophysical phenomena \cite{Kaltenborn:2017hus,Bastian:2017fzo,Fischer:2017lag,Bauswein:2018bma}.
In this work, we restrict ourselves to only the important interactions in order to demonstrate the effect of deconfinement and Mott dissociation
\begin{align}\label{eq:Usfm}
	U^\mathrm{SFM}
	&= D(n_\mathrm{v}) n_\mathrm{q,s}^{2/3}
    	+ a n_\mathrm{q,v}^{2}\,,
\end{align}
with the effective string tension
$D(n_\mathrm v) = D_0 \exp(-\alpha n_\mathrm{q,v}^2)$.
It captures the aspects of (quark) confinement through its resulting density-dependent scalar self-energy contribution which results in an effective suppression of quark \gls{dof} at low densities and temperatures.
Furthermore, we include a minimal description of vector repulsion, which is for the discussion of this paper not relevant, but necessary to move the transition to reasonable densities.

Note that the formulation of \gls{sfm} is based on quark quantities.
To have a consistent description in this paper, we explicitly added an index $\mathrm{q}$ for the quark densities and introduced the scalar quark density $n_\mathrm{q,s} = 3 n_\mathrm{s}$ as well as the vector quark density $n_\mathrm{q,v} = 3 n_\mathrm{v}$, what is important in order to compare the parameters with the original work.

The values of the vacuum string tension $D_0$, the available volume fraction $\alpha$ and the linear vector coupling parameter $a$ are given in the results section.

\subsection{Hadrons}
\label{sec:se:had}
For description of hadronic degrees of freedom we use the \gls{dd2} model \cite{Typel:1999yq,Typel:2009sy}.
It is a nucleonic model with mean field meson interactions of the Walecka type.
The model has proven to be very accurate in describing properties of nuclei and nuclear matter.
In the present work it is only important, that it describes hadronic degrees of freedom and itself already features the nucleonic liquid-gas transition.
The DD2 model can be expressed in the \gls{rdf} formalism with the potential
\begin{align}
	U^\mathrm{DD2} &=
	- \frac 12 \frac{\Gamma_\sigma^2}{m_\sigma^2} n_\mathrm{s}^2
	+ \frac 12 \frac{\Gamma_\omega^2}{m_\omega^2} n_\mathrm v^2
	+ \frac 12 \frac{\Gamma_\rho^2}{m_\rho^2} n_\mathrm{vi}^2\,.
\end{align}
All \gls{dd2} coupling terms and parameters can be found in \cite{Typel:2009sy}.

\section{Results}
\label{sec:results}

For the scenario with a critical endpoint, the parameters are chosen to be $\sqrt{D_0} = \np[MeV]{290}$, $\alpha = \np[fm^6]{0.15}$ and $a = \np[MeV fm^3]{12}$.
This choice is not due to any constraints or physical application.
It just states the most simple example to demonstrate the feature of the cluster expansion in this scenario.

\begin{figure}[tp]
	\centering
	\includegraphics[width=0.48\textwidth,valign=t]{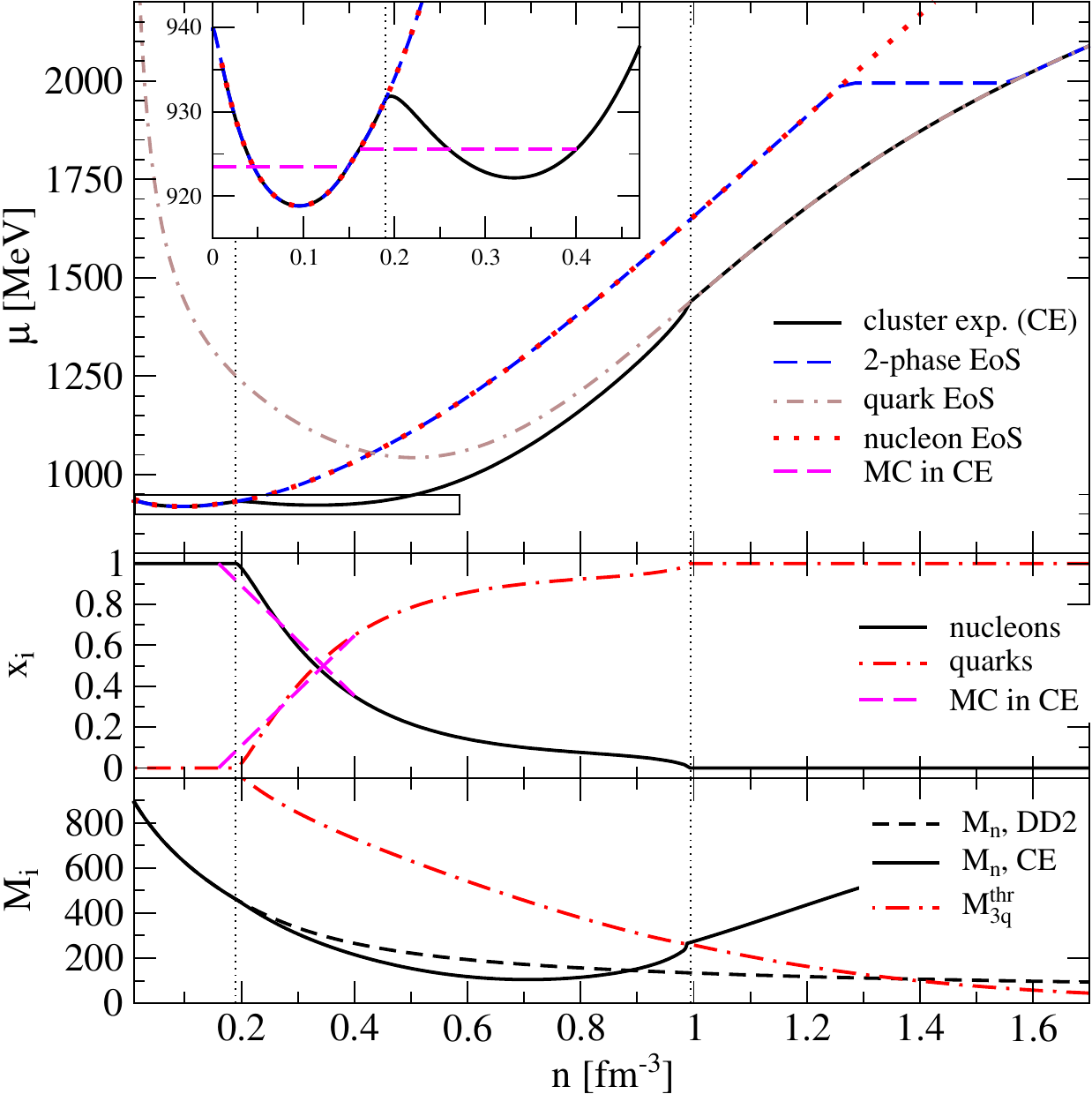}
	\caption{%
	Phase transition from hadronic to quark \gls{eos} comparing the cluster expansion (CE) and the two-phase approach.
	The upper panel shows the chemical potential over density, while the green dashed box is magnified in the inset, showing the phase transition of the CE.
	The middle panel shows the concentration of hadrons and quarks within the CE.
	The lower panel shows the threshold mass and the effective mass of nucleons of our model, compared to the purely nucleonic model.
	The vertical black thin dotted lines show the region, where both quarks and hadrons exist in the system for the CE and the dashed magenta lines are modifications of the Maxwell construction.
	}
	\label{fig:mu_n}
\end{figure}

The resulting \gls{eos} can be seen in \cref{fig:mu_n}, where we compare the common two-phase approach with the cluster-virial expansion, utilizing the Maxwell construction in both cases.
Once there is an occupation of quarks in the system, the \gls{eos} gets altered from the pure hadronic one.
In our case this leads directly into thermodynamic instability ${\partial\mu}/{\partial n} < 0$, most likely due to choice of parameters.
This instability can be seen as a second Van-der-Waals wiggle and hence indicates a first-order phase transition as it can be seen in the upper panel of \cref{fig:mu_n}.
Now we have two first-order phase transitions, one for the pure hadronic liquid-gas transition and one for the quark-hadron transition, connected to the deconfinement of quarks.
Due to the extended persistence of hadrons as bound states in the quark-gluon plasma, this phase gets softer and the phase transition is at much lower densities than in the two-phase approach.

In the lower panel we show the threshold mass $M^\mathrm{thr}_\mathrm n$ and the effective mass of nucleons $M_\mathrm n$, which depends linearly on the scalar density $n_\mathrm{s}$ of the system.
Once quarks contribute to $n_\mathrm{s}$, $M_\mathrm n$ gets lowered, compared to the model without quarks.
The reduction of nucleons in the system has a stronger effects to $n_\mathrm{s}$ than the increase of quarks, which causes an extremum in $n_\mathrm{s}$ and $M_\mathrm n$.
This effect is not seen in quarks, because of the vector-density-dependent effective string tension, see \cref{eq:Usfm}.

\begin{figure}[tp]
	\centering
	\includegraphics[scale=\agrscale,valign=t]{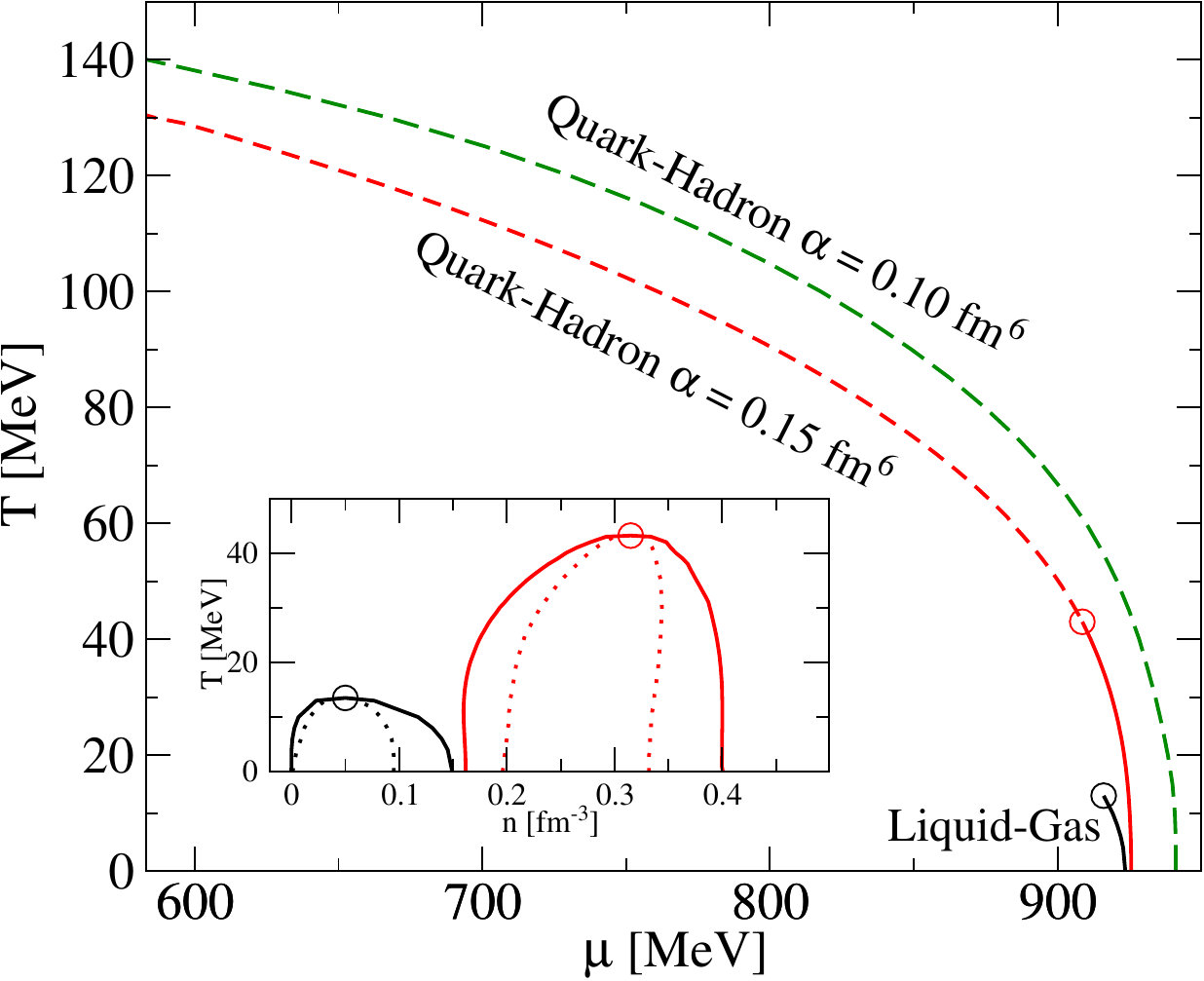}
	\caption{%
	Phase diagram of the cluster expansion, showing temperature over baryon chemical potential.
    The well known liquid-gas phase transition (black) and two parametrizations of quark-hadron transition are displayed.
    One features a first order phase transition (red) with mixed phase and a critical endpoint at $T_\mathrm C = \np[MeV]{43}$ and $\mu_\mathrm C = \np[MeV]{909}$.
	The other represents a crossover all over scenario (green) without critical endpoint.
    The dashed lines show the crossover transition, which is retrieved from the inflection point in chemical potential
    over density.
    The inset shows the temperature over baryon density, including mixed phases for the case with critical endpoint.
    The borders of the phase coexistence regions, the binodals, for the liquid-gas (black) and quark-hadron (red) transitions are shown by the solid lines resulting from a Maxwell constructions.
    The dotted lines show the spinodal region of thermodynamic instability (${\partial\mu}/{\partial n} < 0$).
    }
	\label{fig:phasediagram}
\end{figure}

In \cref{fig:phasediagram} we show a phase diagram, based on Maxwell constructions.
Besides the liquid-gas transition, the quark-hadron phase transition has a critical endpoint, which separates the region of first-order transitions from that of smooth crossover transitions.
The crossover was obtained from the inflection point in the $\mu-n$ plot.

This calculation features a very tiny region of hadronic liquid and the critical temperature of the quark-hadron transition is comparably small.
Once again we want to mention, that the parameter choice is purely academic and a thorough parameter scan needs to be performed.

An interesting outcome of our model is the fact that the thermodynamic phase transition does not coincide with the microscopic Mott transition (nucleon dissociation).
Besides the liquid-gas transition ($n_\mathrm B < \np[fm^{-3}]{0.15}$), we obtain 4 different regions along the density axis (in brackets we indicate the borders at $T=0$ as it is shown in \cref{fig:mu_n}):
\begin{enumerate}
	\item Homogeneous hadron liquid
		\\($\np[fm^{-3}]{0.15} < n_\mathrm B < \np[fm^{-3}]{0.16}$)
	\item Binodal region for the coexistence of hadron liquid and QGP, which we model with a Maxwell construction which assuming a sufficiently large surface tension between the phases \cite{Voskresensky:2001jq}
    	\\($\np[fm^{-3}]{0.16} < n_\mathrm B < \np[fm^{-3}]{0.4}$)\\
which encloses a spinodal region of mechanical instability        \\($\np[fm^{-3}]{0.19} < n_\mathrm B < \np[fm^{-3}]{0.33}$)
	\item Homogeneous multi-component plasma with quarks and remaining bound hadrons
    	\\($\np[fm^{-3}]{0.4} < n_\mathrm B < \np[fm^{-3}]{0.99}$)
	\item Homogeneous pure quark plasma
    	\\($\np[fm^{-3}]{0.99} < n_\mathrm B$)
\end{enumerate}
The appearance of quarks at $n_\mathrm B = \np[fm^{-3}]{0.19}$ is triggering the thermodynamic instability (spinodal).
The existence of the third phase is a crucial difference to the two-phase approach.
This phase is on the high-density end of the Maxwell construction.
Due to the additional (hadronic) degrees of freedom it is much softer (compared to the pure quark phase), resulting in the exceptionally early onset.

Another parameter set, featuring the possibility of a crossover-all-over scenario, one gets by changing the volume fraction parameter to $\alpha = \np[fm^6]{0.10}$, which controls how fast deconfinement is reached along density \cite{Kaltenborn:2017hus}.
The corresponding line of the crossover in the phase diagram can be seen in \cref{fig:phasediagram} as well.
With these parameters, the two-phase approach would still give a first-order phase transition, at slightly higher densities.

Even though one can see in the inset of \cref{fig:phasediagram} a slight increase of the spinodal region at moderate temperatures, there was no scenario found with two critical endpoints.
Such scenario is suggested as a possibility of the quark-hadron continuity \cite{Schafer:1998ef,Hatsuda:2006ps,Hirono:2018fjr}
due to the appearance of non-vanishing diquark condensate signalling a colour-superconducting phase of quark matter.
The inclusion of the diquark correlations is straightforward in the present approach but goes beyond the scope of this exploratory work.

\section{Conclusions}
\label{sec:conclusions}
We have demonstrated that a unified quark-nuclear \gls{eos} can be obtained within a cluster expansion for strongly correlated quark matter, where the clusters are baryons with spectral properties that are described within the generalized Beth-Uhlenbeck approach by a medium--dependent phase shift.
As an instructive example we have discussed here the simple and generic model for the phase shift of baryons with a step-up at the effective mass of the baryon describing an on-shell bound state and a step-down located at the mass of the three-quark continuum threshold which models the continuum as an anti-bound state.
This simple ansatz fulfils the Levinson theorem by construction.
The quark and baryon interactions are accounted for by the coupling to scalar and vector meson mean fields modelled by density functionals.
At increasing density and temperature, due to the different medium dependence of quark and baryon masses, the Mott dissociation of baryons occurs and the contributions of nuclear clusters to the thermodynamics vanish.
It is demonstrated on this simple example that this unified approach to quark-nuclear matter is capable of describing a first-order phase transition with a critical endpoint as well as the case of a crossover-all-over.

The next step would be to include a larger set of particles like diquarks, mesons and baryons and to include a more sophisticated functional for the quark interaction. 
With a sufficient amount of particles it should be possible to adjust the parameter set in such a way that it follows Lattice QCD calculations at zero density and fulfils known constraints for low temperatures.
Furthermore, a generalization to arbitrary charge-fraction (isospin-asymmetry) is straightforward and would make the model applicable for astrophysical systems.

\section*{Acknowledgements}
The authors thank Gerd R\"opke and Elizaveta Nazarova for critical reading of the manuscript and discussions.
The work of N.-U.F.B. was supported by the Polish National Science Center (NCN) under contract number 2019/32/C/ST2/00556.
The work of D.B. was supported by the Russian Science Foundation under grant number 17-12-01427, by the NCN under contract number 2019/33/B/ST9/03059 and by the National Research Nuclear University (MEPhI) within the Russian Academic Excellence Project under contract number 02.a03.21.0005. 
The authors thank the COST Actions CA15213 ``THOR'' and CA16214 ``PHAROS'' for supporting their networking activities.

\bibliography{literature_notitle}

\end{document}